\documentstyle[preprint,aps,epsf]{revtex}

\newcommand{\beq}{\begin{eqnarray}}
\newcommand{\eeq}{\end{eqnarray}}

\newcommand{\btem}{\bibitem}

\begin{document}

\preprint{UTHEP-353, RYUTHP-96/3, Dec. 1996}

\draft

\title{Optimized Perturbation Theory\\  for Wave Functions of Quantum 
 Systems }

\author{T. Hatsuda, T. Kunihiro,$^1$
 and  T. Tanaka}

\address{Institute of Physics, University of Tsukuba,
 Tsukuba, Ibaraki 305, Japan}

\address{$^1$Faculty of Science and Technology, Ryukoku University,
 Seta, Ohtsu, 520-21, Japan}

 
\maketitle

\begin{abstract}
The notion of the  optimized perturbation, 
which has been successfully applied to  energy eigenvalues, is  generalized
 to treat  wave functions of quantum systems. 
 The key ingredient is to
construct an envelope of a set of 
 perturbative wave functions. This leads to a condition
 similar to that obtained from the principle of minimal sensitivity. 
 Applications of the method 
 to quantum anharmonic oscillator and the double well potential show that 
  uniformly valid wave functions  with correct asymptotic behavior
 are obtained in the first-order optimized 
 perturbation even for strong couplings.

\end{abstract}

\narrowtext
Naive perturbative expansions are usually  divergent asymptotic
 series \cite{rev1}.
 Thus various summation techniques  such as the Borel and 
  Pad\'{e} methods  have been developed  \cite{rev2}. 
  In recent years,  a new summation method
  ``optimized perturbation theory (OPT)'' has been studied.
  This combines straightforward perturbation with  
  a variational principle and  improves 
 even the Borel non-summable series.
 Variants of OPT are called in various names,
the (optimized) $\delta$-expansion,
  the variational perturbation, the renormalized strong 
 coupling expansion,  and so on \cite{delta}. 
 One of the key ingredients in the method is 
 the principle of minimal sensitivity
 (PMS) \cite{PMS},
 which says that physical quantities calculated in perturbation theory 
  should not depend on any parameters absent in the original 
 Hamiltonian. 

  The method has been applied to improve the energy eigenvalues of the
 quantum anharmonic oscillator (AHO) and the double-well potential (DWP);
  AHO and DWP  are 
 the useful laboratories to test a 
 new non-perturbative method. A rigorous proof has been also given 
 for the  convergence properties  of the perturbation series 
 for energy eigenvalues of AHO/DWP \cite{proof}.
 The method is also being extensively applied to
 perturbation series for energy and partition functions 
 in quantum field theories \cite{appl}.

  To the best of our knowledge, the wave functions
 has never been explored in OPT  even for simple quantum  mechanical problems.
 Needless to say, wave functions have all information of the system in 
question, including the energy eigenvalues. 
 The purpose of this Letter is to show a novel
 generalization of  the idea of the optimized
 perturbation  for  studying the wave functions.

 Let's take the following Schr\"{o}dinger equation for one dimensional 
AHO/DWP;
\beq
\label{ham1}
{1 \over 2} (-  {d^2 \over d x^2}
 \pm x^2 + \lambda x^4)
\psi_n (x) = E_n (\lambda) \psi_n(x). 
\eeq
 Naive  Rayleigh-Schr\"{o}dinger (RS) expansion for eigenvalues
 $E(\lambda ) = \sum_n \lambda^n C_n$ gives an asymptotic series
with coefficients 
 growing as $C_n \sim n ! $ \cite{BW}.
 A scaled Hamiltonian in the simplest version of OPT
 (sometimes called the linear $\delta$-expansion) reads
 $H_{\delta} = H_0(\Omega) + \delta \cdot H_I (\Omega, \lambda)$ with
\beq
\label{ham2}
H_0 (\Omega) & = & {1 \over 2} ( -  {d^2 \over d x^2} + \Omega^2 x^2 ) , \\
\label{ham3}
H_I (\Omega, \lambda) & = & {1 \over 2} \left( 
(\pm 1- \Omega^2)x^2 + \lambda x^4 \right) .
 \eeq
$H_{\delta}$ interpolate the free Hamiltonian with a frequency $\Omega$
  and the full Hamiltonian; $H_{\delta =0} =H_0(\Omega)$ and
 $H_{\delta =1} = H$.
   The essential idea of OPT is to choose
 a suitable $\Omega$ so that the RS perturbation 
  $E(\lambda) = \sum_n \delta^n C_n $ 
 becomes a  convergent series  with errors exponentially
 suppressed for large $n$.
   PMS in this case is one of the 
 possible criteria for choosing $\Omega$,
  although it is not a unique one \cite{PMS}.

 To show how the naive 
 application of the idea of OPT to the wave function fails,
 we first  expand the wave function of $H_{\delta}$ by
the eigenfunctions of $H_0$,
\beq
\label{RS}
 \psi_k(x;\Omega) = \sum_{j} C_{jk} (\Omega) \psi_j^0 (x; \Omega).
\eeq
 $C_{jk}(\Omega)$ is evaluated in  the RS perturbation theory with 
$\delta$ being  an expansion parameter, 
$C_{jk} (\Omega) = C_{jk}^0 + \delta\cdot C_{jk}^1 + \cdot \cdot  \cdot $.
 With a suitable $\Omega$, one might expect 
  $\psi(x,\Omega)$ at $\delta =1$ converges rapidly
 to the exact wave function $\psi^{\rm exact}(x)$ as in the case of
 the energy eigenvalues $E$.
  However, this is not the case as long as $\Omega$ is  independent
 of $x$. In fact, 
  in any finite orders of the expansion in (\ref{RS}),
  $\psi_n(x \rightarrow \infty ;\Omega) \sim \exp(-\Omega x^2/2)$, while
 the exact wave function in (\ref{ham1}) behaves as
\beq
\label{asymp}
\psi^{\rm exact}
 (x \rightarrow \infty) \sim \exp(- \sqrt{\lambda \over 9}\mid x \mid ^3).
\eeq
 Thus, the series (\ref{RS}) can never reproduce the 
 true  asymptotic behavior no matter what $\Omega$ is chosen as long as
 $\Omega$ is constant.

 To find a way out, let us  describe the situation in geometrical  terms.
 The perturbative wave function  (\ref{RS})  
 with constant $\Omega$ gives a locally valid approximation around
 certain $x$ where the perturbation $\delta \cdot H_I (\Omega)$
 is sufficiently small; 
 however, this is not a uniformly valid solution in the entire domain
 of $x$.
 Conversely speaking, if $\Omega$ is varied,
   we have a family of functions $\{\psi(x;\Omega)\}_{\Omega}$ 
 parametrized with $\Omega$, each of which is locally valid around
 certain $x$. 
 Therefore  a uniformly valid wave function may be constructed as an  
 envelope of the family of perturbative ones.
  To obtain an equation to determine the envelope, 
  we   recall the classical theory of envelopes
 \cite{env0}: Suppose one has a set of functions $f(x;\alpha)$
 parametrized by  $\alpha$.
  Then the envelope of $\{ f(x;\alpha ) \} _{\alpha}$  can be 
 obtained by solving 
 $\partial f(x;\alpha) /\partial \alpha =0 $ and substituting 
 the solution $\alpha(x)$ into the original
 function i.e. $f(x;\alpha(x))$. In our case, 
 $f=\psi$ and $\alpha = \Omega$. Thus we reach a condition,
\beq
\label{PMS}
 {\partial \psi(x; \Omega)  \over \partial \Omega }\Biggl\vert
 _{\delta =1} =0. 
\eeq
 This condition has another interpretation in term of  PMS:
 The exact wave functions of (\ref{ham1}) do not depend on the 
 fictitious PMS parameter $\Omega$ at $\delta =1$. So one would impose  
  the  condition that even the approximate wave function should not depend
 on $\Omega$, which leads to eq.(\ref{PMS}).
 A similarity of this idea to that of the renormalization group   
  \`{a} la Gell-Mann-Low is also to be noted.

 At this point,  one may ask  whether   
 $\psi(x;\Omega(x))$ thus obtained  satisfies the
 original Schr\"{o}dinger equation where  
 $d/dx$ acts on $\Omega(x)$ too: 
$ (d/dx) \psi(x;\Omega(x))
 = (\partial /\partial x) \psi(x;\Omega(x))  
 +  (\partial \Omega(x)/ \partial x) \cdot
  (\partial / \partial \Omega )  \psi(x;\Omega(x))$.
 The answer is yes, since  the second term 
 vanishes on account of  the condition (\ref{PMS}). 
 The same is true for $d^2/dx^2$.
 Hence the improved wave function satisfies the Schr\"{o}dinger
 equation at every point $x$  at least up to the same order of the 
 $\delta$-expansion employed.

Solution of (\ref{PMS}) gives $\Omega$ as
 a function of  $x$ (and $\lambda$).  For AHO/DWP, eq. (\ref{PMS})
 turns out to be an algebraic one.
 The envelope wave function $\psi(x;\Omega(x))$ 
 obtained by substituting the solution of (\ref{PMS}) into (\ref{RS})
  turns out to be a uniformly valid approximation for
 $\psi^{\rm exact}(x)$  in a global domain 
 even in the lowest order of the $\delta$-expansion.

 Let us demonstrate the above points explicitly
 for  AHO/DWP with the first order $\delta$-expansion in which
  $C_{jk} = \delta_{jk} - 
 \langle j \mid H_I \mid k \rangle /(E_j^0 - E_k^0) $.
  The condition eq. (\ref{PMS}) gives an algebraic 
 equation for $\Omega$ as
\beq
\label{PMS2}
\Omega (\pm 1-\Omega^2)
\left(A_n H_n(\sqrt{\Omega} x)
 + 4n(n-1) A_{n-2} H_{n-2}(\sqrt{\Omega} x) \right) \\
 \nonumber
  + \lambda \cdot 
\left( B_n H_n(\sqrt{\Omega} x) 
+ 4n(n-1) B_{n-2} H_{n-2}(\sqrt{\Omega} x)\right)  =0,
\eeq 
 where $H_n(z)$ is the n-th Hermite polynomial, and the 
 functions  $A_n(\sqrt{\Omega} x)$ and $B_n(\sqrt{\Omega} x)$ are
 given as follows;
 $ A_n(z) = 16 z^4 - 16 (2n-1) z^2 - 4(8n+7),$
 $A_{n-2}(z)   = -16$,
 $B_n(z) = 8z^6 + 4(2n+9)z^4 -2(24n^2+4n-33)z^2-9(8n^2+26n+11)$, 
 and $B_{n-2}(z)   = -2(8z^2 +18n +9)$.

Several remarks on  
Eq.(\ref{PMS2}) are in order here.

\noindent
1. Eq.(\ref{PMS2}) gives an algebraic equation of degree
 $m+5$ for $\Omega $ with $n=2m$ or $2m+1$.
 As in the usual PMS condition applied to the energy eigenvalues,
 one needs some principle to  select a relevant  root.  
 The true solution $\Omega(x)$  is selected
 so that the local energy ${\cal E}_n(x)$ defined by
 $\psi(x;\Omega(x))^{-1}\ H\ \psi(x;\Omega(x)) $ is 
 (roughly)  constant as a function of  $x$.

\noindent
2. For
$\lambda=0$, there is an obvious solution $\Omega(x) =1$ for (\ref{PMS2}).
 Other solutions do not satisfies ${\cal E}_n(x)\simeq$constant.

\noindent
3.  For $x \rightarrow \infty $, eq.(\ref{PMS2}) with an
 ansatz $\Omega(x) \sim C x^p (p > 0)$ gives 
 an equation independent of $n$. It has a solution
 $C^2=\lambda /2$ and $p=1$.
Thus the asymptotic form  reads
\beq
\label{asymp2}
\psi(x;\Omega(x)) \rightarrow \exp (- \sqrt{\lambda \over 8} \mid x\mid^3 )
\eeq
which has a same asymptotic behavior with  (\ref{asymp}).
This feature is never obtained by any finite-order RS
  perturbation as we mentioned. 
 A slightly different numerical coefficient of  $\vert x\vert^3$
  is to be improved with  higher order $\delta$-expansion \cite{bend}.

\noindent
4.
 Although $\psi(x;\Omega)$  with constant $\Omega$ is normalized as
 $\int_{-\infty}^{ \infty} \mid \psi(x;\Omega) \mid^2 dx =1$, the 
 norm of  $\psi(x;\Omega(x))$ deviates
 from unity due to the $x$ dependence of $\Omega$. In the following figures,
 we will always show  renormalized wave functions.

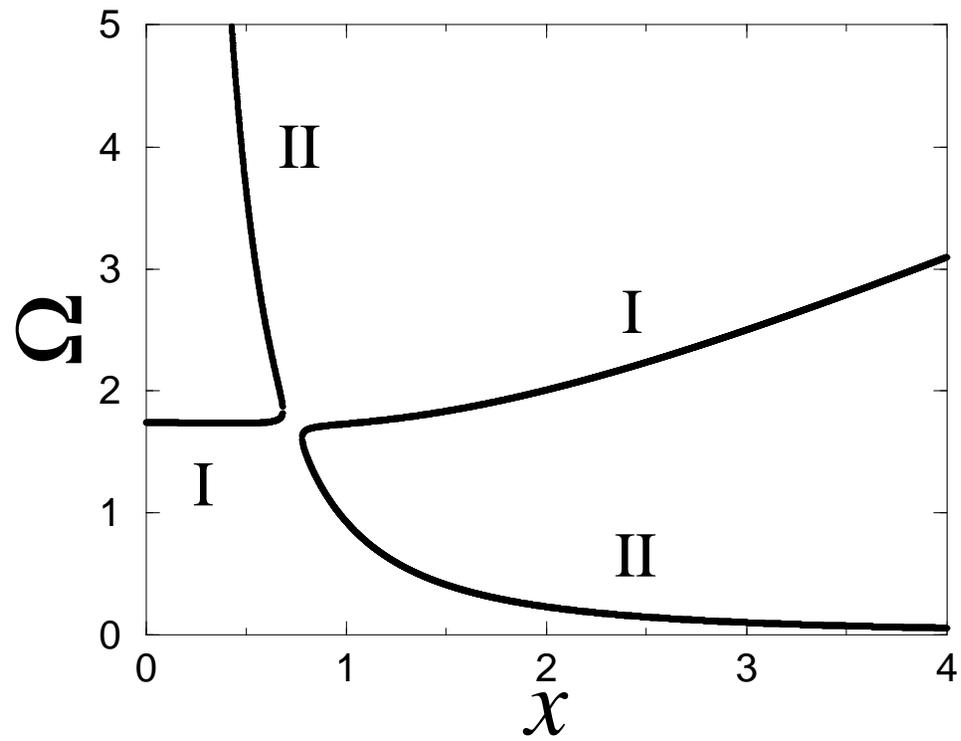
\begin{figure}
\caption{$\Omega(x)$ as roots of the algebraic equation
for the ground state with intermediate coupling $\lambda =1$.
  The branch I  gives approximately constant local-energy.}
\label{fig:fig1} 
\end{figure}

 In fig.~\ref{fig:fig1},  the real and positive 
 solutions of (\ref{PMS2}) for
 the ground state of AHO with intermediate coupling ($\lambda =1$)
 are shown as a function of $x$. 
  The local energy is approximately constant for the branch I.
 In the interval  $ 0.684 \leq x \leq  0.780$,   
  $\Omega(x)$ becomes complex with a small
 imaginary part. When we construct a continuous wave function in this
 interval,
  we adopt a real part of $\Omega(x)$ as a simplest way of 
  interpolation\cite{comm3}.
  We expect that the width of this interval and the associated 
  small imaginary part of $\Omega(x)$ 
  will vanish  in the higher order $\delta$-expansion.
 This must be, however, checked explicitly, which is now under progress.

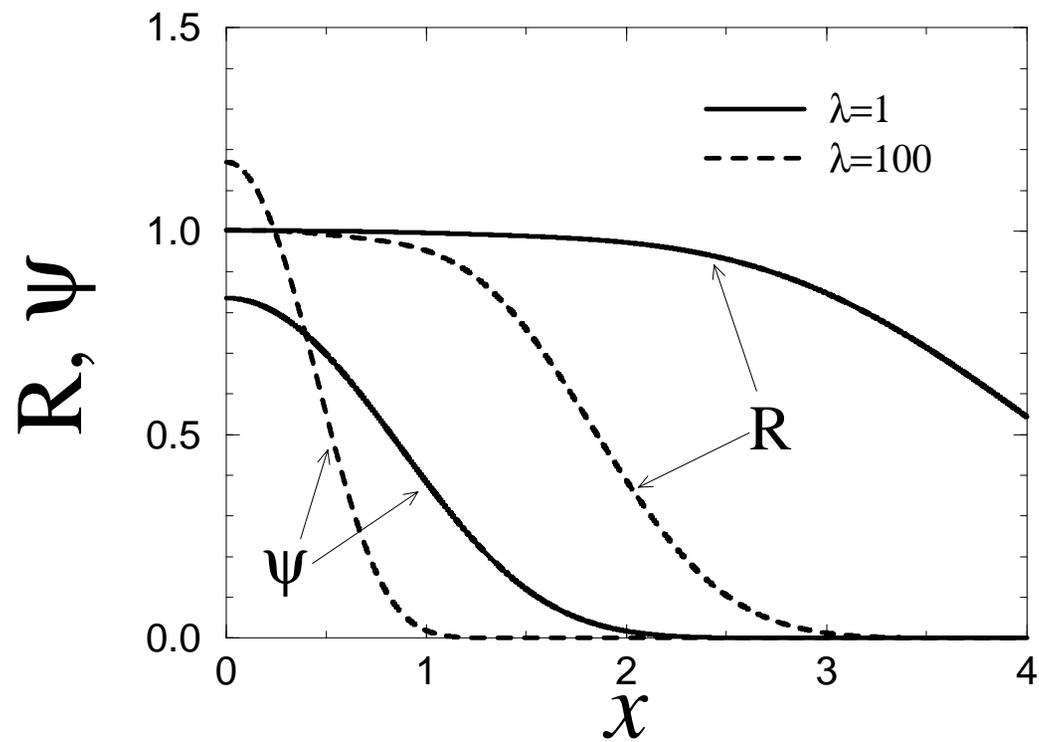
\begin{figure}
\caption{The ratio 
 $R \equiv \psi(x;\Omega(x))/\psi^{\rm exact}(x)$ 
 as well as $\psi(x;\Omega(x))$ itself for the
 ground state.
The solid (dashed) line is for the intermediate (strong)
 coupling $\lambda =1 (100)$. }
\label{fig:fig2}
\end{figure}

In fig. ~\ref{fig:fig2}, the ratio 
 $R \equiv \psi(x;\Omega(x))/\psi^{\rm exact}(x)$ as well as
  $\psi(x;\Omega(x))$ for the ground state of AHO are shown
  for intermediate coupling ($\lambda =1$) and strong coupling
 ($\lambda =100$).  $\psi^{\rm exact}(x)$
 is numerically solved in double precision
 using the shooting method with FORTRAN \cite{comment2}.

It is not until $x$ exceeds $x_c = 2.725 (1.211)$ for $\lambda =1 (100)$ 
 that the relative error of $\psi(x;\Omega(x))$ to the 
 exact one  becomes  more than $10\%$.
 Such a value of $x_c$ is large enough in the sense that 
 the absolute value of the wave function at $x_c$ is already 
 as tiny as $\psi(x_c) = 0.00016 (0.0012)$. Hence
 our first order
 results can be used as  an exact one for practical use of the
 wave function.
 In fig.~\ref{fig:fig2}, $O(1)$ deviation is seen for quite  
 large values of $x$.
 This is due to the approximate asymptotic behavior of 
 $\psi(x;\Omega(x))$. The ratio $R$ for large $x$ can be estimated from
 (\ref{asymp}) and (\ref{asymp2}) as
 $R \sim \exp(- 0.02 \sqrt{\lambda} \mid x \mid ^3)$.  

To see the accuracy of the wave function  more quantitatively, 
 a comparison of $\psi^{\rm exact}(x)$ and $\psi(x; \Omega (x))$  
 for $\lambda =100$  
 is shown in Table \ref{tab:tab1} for the ground state of AHO.

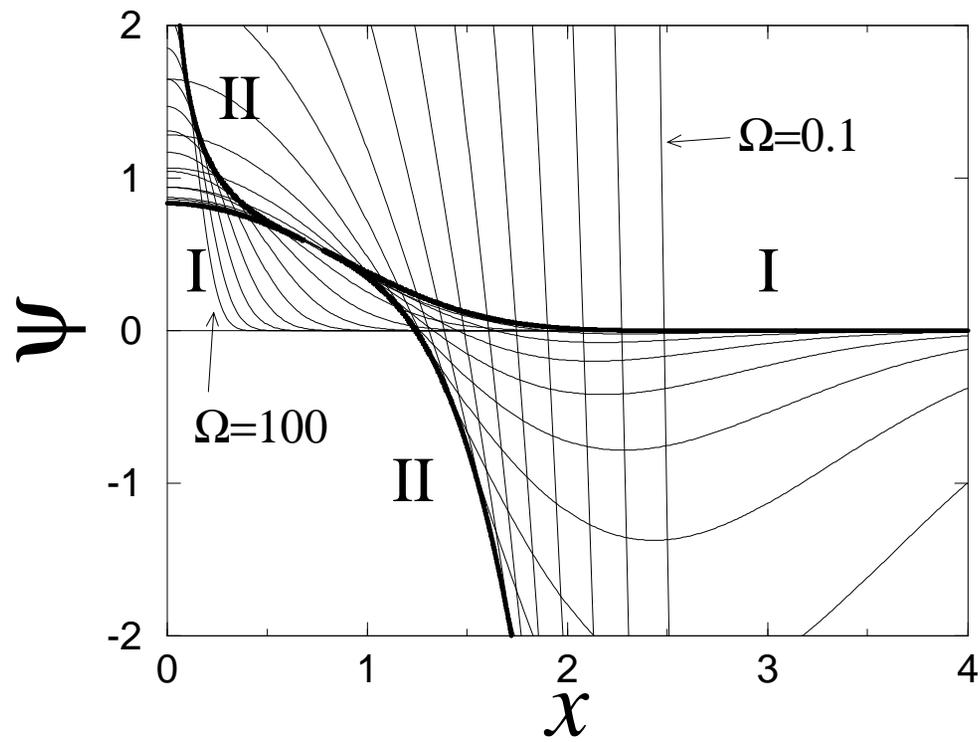
\begin{figure}
\caption{Family of perturbative wave functions with constant $\Omega$
 with $0.1 \leq \Omega \leq 100$ for the ground state with $\lambda =1$.
Thick lines labeled as I and II are their envelops.
}
\label{fig:fig3}
\end{figure}

 As we have mentioned before, $\psi(x;\Omega(x))$ 
 is the  envelop of the perturbative wave functions
 parametrized by $\Omega$.  In fig.~\ref{fig:fig3},
 thin lines are the wave functions for $\lambda =1$
 with constant $\Omega$ varied.
 There are two envelopes of this family of wave functions, which are
 labeled by I and II in the figure.
 $\psi(x;\Omega(x))$  is obtained by the envelop I.

In Table \ref{tab:tab2},  expectation values of 
 several operators including the full Hamiltonian $H$ 
 with respect to  $\psi(x;\Omega(x))$  are compared with the exact numbers.
 The agreement is excellent as a first-order approximation.

Some  final remarks are in order here.

\noindent
1. Although we have shown only the ground state wave function because of the
 limitation of space, we have also studied the excited states ($n \neq 0$)
 and found that $\psi(x;\Omega(x))$ 
  are as good approximation for $\psi^{\rm exact}(x)$ 
  as the $n=0$ case.

\noindent
2. The present method here 
 can be used to study the double well potential where
 the sign of the $x^2$ term in (\ref{ham1}) is negative.
 In the first order $\delta$-expansion, we have found that the 
 exact wave function is accurately reproduced
 for strong couplings, but the approximation breaks down
 qualitatively for $\lambda < 0.065$.
 In such weak couplings, or equivalently  a high barrier at the origin,
 the ground state wave function 
 is well localized in the two degenerate minima of the
 double well potential. Therefore, it is natural to expect that  
the naive $\delta$-expansion scheme
 using $H_0 (\Omega)$ with a single harmonic potential
 does not work.  An introduction of a new PMS parameter
  is under investigation to widen the applicability of the
 method.

\noindent
3.  
 We have also performed 
the second order $\delta$-expansion for AHO.
 The formula is more lengthy and the degree of the algebraic equation
 is higher.  The detailed comparison with the first order
 result is under progress.

\noindent
4. 
 Applications to other quantum mechanical system such as
 the scattering problem, Zeeman effect, and Stark effect \cite{rev1}  
 are interesting. Also, applications to field theory such as  
  the improvement of the space-time  (or momentum) dependence
 of  Green's functions are closely related to the
 problem studied in this work.

\noindent
5.  The notion of envelopes  has been found useful 
 for  global asymptotic analysis of some of 
ordinary/partial differential equations having {\em secular terms} 
in perturbative solutions\cite{kuni}. 
The present work together with those in 
\cite{kuni}  shows that the notion of  envelopes 
 can be   a  useful guide for the improvement of
   perturbative expansions and asymptotic analyses.\cite{comm5}


In summary, we have developed an optimized perturbation theory for
 wave functions for AHO and DWP.
  The optimized wave functions are constructed
 as an envelope of perturbative ones, and 
 the principle of minimal sensitivity can be
 regarded as a condition for constructing envelopes.
  The optimized wave function  thus obtained in the
 first-order $\delta$-expansion has a correct
 asymptotic behavior at $x \rightarrow \infty$ and 
 is uniformly valid for entire domain of $x$ 
 even for the  strong coupling.

\section*{Acknowledgements}

 The authors thank H. Kitagawa for his help of solving 
 the Schr\"{o}dinger equation numerically in high accuracy.
 This work was supported  by 
  the Grants-in-Aid of the Japanese Ministry of 
Education, Science and Culture (No. 06102004), (No. 07304065) and 
(No. 08640396).

\newpage

\begin{table}  
\begin{center}
\begin{tabular}{|r|l|l|l|l|l|}
 & $x =0.0$ &  $x=0.5$  &  $x=1.0$  &  $x=1.5$  &  $x=2.0$
 \\ \hline \hline 
 $\psi(x;\Omega(x))$    
	& 1.170E0 & 5.519E-1 & 1.885E-2 & 3.925E-6 & 3.095E-13 \\ 
 $\psi^{\rm exact}(x)$  
	& 1.167E0 & 5.564E-1 & 1.979E-2 & 5.168E-6 & 7.982E-13 \\
\end{tabular}
\end{center}
\caption{Ground state wave functions of AHO for $\lambda = 100$;
 $\psi(x;\Omega(x))$ is an approximate wave function in the 
 first order $\delta$-expansion.}
\label{tab:tab1}
\end{table}

\begin{table}  
\begin{center}
\begin{tabular}{|r|c|c|c|c|}
 & $\langle p^2 \rangle $ & $\langle x^2 \rangle $ & $\langle x^4 \rangle $
 & $\langle H \rangle $  \\ \hline \hline
 $\lambda =1$ \ \ \ \ PMS   
	& 0.83146 & 0.30391 & 0.25703 & 0.69620 \\
              \ \ \ \ Exact 
	& 0.82630 & 0.30581 & 0.26024 & 0.69618 \\ \hline
$\lambda = 100$ \ \ \ PMS   
	& 3.34430 & 0.07641 & 0.01579 & 2.50003 \\
              \ \ \ \ Exact 
	& 3.30717 & 0.07731 & 0.01615 & 2.49971 \\ 
\end{tabular}
\end{center}
\caption{Expectation values 
 of $p^2 =(-id/dx)^2$, $x^2$, $x^4$ and the full Hamiltonian $H$
 for the ground state in AHO.}
\label{tab:tab2}
\end{table}


\begin{references}

\btem{rev1}  {\em Large Order Bahavior of Perturbation Theory},
 Current Physics -- Sources and Comments,  vol.7,
 ed. J. C. Le Guillou and J. Zinn-Justin, (North-Holland, Amsterdam, 1990).

\btem{rev2} G. A. Arteca, F. M. Fern\'{a}ndez and E. A. Castro, 
{\em Large Order Perturbation Theory and Summation Methods
 in Quantum Mechanics}, (Springer-Verlag, Berlin, 1990).

\btem{delta} 
We list only the recent articles here from which one can
 trace the enormous number of references related to this method;
 C. Arvanitis, H. F. Jones and C. S. Parker, Phys. Rev. {\bf D52}, 3704
 (1995).
 W. Janke and H. Kleinert, Phys. Rev. Lett. {\bf 75}.
 2787 (1995). E. J. Weniger, Phys. Rev. Lett. {\bf 77}, (1996) 2859.
 H. Kleinert, {\em Path Integrals in Quantum Mechanics, Statistical and 
Polymer Physics}, 2nd. edition (World Scientific, Singapore, 1995).

\btem{PMS} 
 P. M. Stevenson, Phys. Rev. {\bf D23}, 2916 (1981).

\btem{proof} A. Duncan and H. F. Jones, Phys. Rev. {\bf D47}, 2560 (1993).
 C. Arvanitis, H. F. Jones and C. S. Parker, Phys. Rev. {\bf D52},
 3704 (1995). R. Guida, K. Konishi and H. Suzuki, Ann. Phys. {\bf 241},
  152 (1995); ibid. {\bf 249}, 109 (1996).

\btem{appl} D. Gromes, Z. Phys. {\bf C71}, 347 (1996).  A. A. Penin and
 A. A. Pivovarov, Phys. Lett. {\bf B367}, 342 (1996).
 G. Krein, D. P. Menezes and M. B. Pinto, Phys. Lett. {\bf B370}, 5 (1996).


\btem{BW} 
 C. M. Bender and T. T. Wu, Phys. Rev. {\bf 184}, 1231 (1969); ibid.
 {\bf D7},  1620 (1973).

\btem{env0}
 See any textbooks on calculus.


\btem{bend}
 A multiple-scale perturbation theory, which is a
 powerful method  for asymptotic analysis, has been recently
  applied to study
 the eigenvalues and eigenfunctions of AHO;
 C. M. Bender and L. M. A.
 Bettencourt, Phys. Rev. Lett. {\bf 77}, 4114 (1996), {\tt hep-th/9607074}. 
 They obtain the asymptotic form of the wave function (\ref{asymp})
  only after further resummation.
 Relation between our approach and theirs is not  clear. 

\btem{kuni}
 T.Kunihiro, Prog. Theor. Phys. {\bf 94}, 503 
(1995); (E) {\bf 95}, 835
  (1996);\ Jpn. J. Ind. Appl. Math. {\bf 14} (1997); {\tt hep-th/9609045}.
 Most of equations treated there was first examined by 
the renormalization group  method;
 L.-Y. Chen, N. Goldenfeld and Y. Oono, Phys. Rev. Lett.
 {\bf 73}, 1311 (1994).

\btem{comm3} We stress that the asymptotic behavior (\ref{asymp}) is 
  independent of the interpolation method adopted.

\btem{comment2}
  We have checked
 that our exact eigenvalues numerically obtained 
 agree  up to 12 digits with the exact eigenvalues 
  for $\lambda =1, 100$ given by F. Vinette and 
 J. \v{C}\'{i}\v{z}ek, J. Math. Phys. {\bf 32}, 3392 (1991).
 Also,  $\psi^{\rm exact}(x)$ 
 numerically solved for $\lambda =0$ agrees to the analytic answer
 $\pi^{-1/4} \exp (-x^2/2)$
 up to 10 digits in the interval $0 \leq x \leq 9$.
  Thus our numerical results for $\psi^{\rm exact}(x)$ 
  are accurate enough to be reference values.
 
\btem{comm5} The improved energy eigenvalue obtained by PMS in  previous
 works may be
  interpreted as an  envelope $E(\lambda; \Omega(\lambda))$
  of a set of perturbative energy eigenvalues 
  $\{ E(\lambda; \Omega) \}_{\Omega}$.





\end{references}
\end{document}